\RequirePackage[]{lineno} 
\documentclass[prb,aps,reprint,superscriptaddress,longbibliography]{revtex4-1}

\usepackage{graphicx,epsfig,float}
\usepackage{dcolumn}
\usepackage{bm}    
\usepackage{amssymb} 
\usepackage{amsmath}
\usepackage{verbatim}

\let\oldequation\equation
\let\oldendequation\endequation

\renewenvironment{equation}
  {\linenomathNonumbers\oldequation}
  {\oldendequation\endlinenomath}
\begin{document}


\title{Orbital magneto-optical response of periodic insulators from first principles}

\author{Irina V. Lebedeva}
\email{liv\_ira@hotmail.com}
\affiliation{Nano-Bio Spectroscopy Group and ETSF,  Departamento de F\'isica de Materiales, Universidad del Pa\'is Vasco UPV/EHU, 20018 San Sebasti\'an, Spain}
\author{David A. Strubbe}
\affiliation{Department of Physics, School of Natural Sciences, University of California, Merced, CA, 95343, United States}
\author{Ilya V. Tokatly}
\affiliation{Nano-Bio Spectroscopy Group and ETSF,  Departamento de F\'isica de Materiales, Universidad del Pa\'is Vasco UPV/EHU, 20018 San Sebasti\'an, Spain}
\affiliation{Donostia International Physics Center (DIPC), Manuel de Lardizabal 5, 20018 San Sebasti\'an, Spain}
\affiliation{IKERBASQUE, Basque Foundation for Science, 48011 Bilbao, Spain
}
\author{Angel Rubio}
\email{angel.rubio@mpsd.mpg.de}
\affiliation{Nano-Bio Spectroscopy Group and ETSF,  Departamento de F\'isica de Materiales, Universidad del Pa\'is Vasco UPV/EHU, 20018 San Sebasti\'an, Spain}
\affiliation{Max Planck Institute for the Structure and Dynamics of Matter and Center for Free-Electron Laser Science, Luruper Chaussee 149, 22761 Hamburg, Germany
}

\begin{abstract}
Magneto-optical response, i.e. optical response in the presence of a magnetic field, is commonly used for characterization of materials and in optical communications. However, quantum mechanical description of electric and magnetic fields in crystals is not straightforward as the position operator is ill defined. We present a reformulation of the density matrix perturbation theory for time-dependent electromagnetic fields under periodic boundary conditions, which allows us to treat the orbital magneto-optical response of solids at the \textit{ab initio} level. The efficiency of the computational scheme proposed is comparable to standard linear-response calculations of absorption spectra and the results of tests for molecules and solids agree with the available experimental data. A clear signature of the valley Zeeman effect is revealed in the continuum magneto-optical spectrum of a single layer of hexagonal boron nitride. The present formalism opens the path towards the study of magneto-optical effects in strongly driven low-dimensional systems.
\end{abstract}

\maketitle

\section*{Introduction}
Magneto-optical phenomena originating from the loss of symmetry between left and right circularly polarized light in the presence of a magnetic field are widely used for characterization of different kinds of matter~\cite{barron2004, agranovich1991}. Magnetic circular dichroism (MCD) spectra help to assign overlapping bands and give insight into magnetic properties of the ground and excited states. Faraday rotation of the plane of polarization of linearly polarized light serves as a basic operational principle for functional magneto-optical disks and optical isolators~\cite{sugano1999}. Optical excitations in the presence of a magnetic field allow manipulation of valley pseudospin degrees of freedom in two-dimensional monolayers~\cite{srivastava2015,mitioglu2015,wang2015,macneill2015,li2014,tabert2013}. Giant Faraday rotation has been revealed in graphene~\cite{crassee2011} and metal oxide nanosheets~\cite{osada2006}. These advances cultivate the growing interest to development of a gauge-invariant and computationally efficient \textit{ab initio} theory of magneto-optical response. 

While \textit{ab initio} calculations of MCD spectra in molecules can be performed nowadays in a nearly routine fashion~\cite{solheim2008a, solheim2008b, lee2011, seth2008, seth2008a} (as implemented in quantum chemistry codes \cite{Dalton, ADF}), the complete response theory for extended systems is still under development. The reason is that external electromagnetic fields break the translational symmetry of such systems, which in the formal way is expressed through unboundness of the position operator. Though according to the modern theory of polarization \cite{kingsmith1993,vanderbilt1993,resta1994}, the position operator can be replaced by a derivative with respect to the wave vector in responses to electric fields, the description of magnetic fields is more complicated as it introduces vector coupling to electron dynamics and leads to non-perturbative changes in wavefunctions. Three approaches have been considered in literature to deal with these difficulties: (1) taking a long-wavelength limit of an oscillating perturbation~\cite{mauri1996,shi2007}, (2) using the Wannier function formalism~\cite{thonhauser2005,ceresoli2006,malashevich2010,essin2010} or (3) treating perturbations of the one-particle Green function or one-particle density matrix~\cite{essin2010,chen2011,gonze2011}, which are two-point quantities summed up over all occupied bands and having periodic and gauge-invariant counterparts. While wave functions in the presence of even a very small magnetic field differ drastically from those in the absence of the magnetic field (a plane wave for a free electron and a localized Landau level state for an electron in the magnetic field can be considered as an example), the gauge-invariant counterpart of the density matrix changes perturbatively \cite{essin2010,chen2011,gonze2011}. In approach (1), proper sum rules \cite{sangalli2017,raimbault2015} should be taken into account to control numerical errors arising upon summing up non-gauge-invariant paramagnetic and diamagnetic terms. In approach (3), such a numerical noise is supressed automatically. Approach (3) also allows us to work under purely periodic boundary conditions as opposed to approach (2), where contributions of open boundaries should be treated carefully \cite{thonhauser2005,ceresoli2006,malashevich2010}. 

So far the magnetic field has been considered in the context of static responses \cite{mauri1996,shi2007,thonhauser2005,ceresoli2006,malashevich2010,essin2010,essin2010,chen2011,gonze2011}. In the present paper we demonstrate that density matrix perturbation theory \cite{lazzeri2003,essin2010,gonze2011} can be extended to the case of dynamic non-linear phenomena. We focus on second-order magneto-optical effects, i.e. the change of the optical response in the presence of a magnetic field. While the approach developed here is general and can be adapted to any first-principles framework, we decide to illustrate it using time-dependent density functional theory (TDDFT) \cite{runge1984,marques2012}. This method provides a satisfactory level of accuracy at a moderate computational cost and has been widely employed in literature for magneto-optical response of molecules \cite{solheim2008a, solheim2008b, lee2011, seth2008, seth2008a}. The account of excitonic effects in the transverse optical response of solids, however, is not straightforward within TDDFT and is performed here using the approach 
derived in Ref. \onlinecite{berger2015} from time-dependent current density functional theory (TDCDFT).

The procedures for solids implemented for the present paper form a part of the open-source  code Octopus~\cite{marques2003,castro2006,andrade2015}. For the sake of simplicity, we limit our consideration to orbital magneto-optical effects for insulators. While the spin contribution is trivial, the account of the Fermi surface contribution can be done for metals by analogy with Ref. \onlinecite{shi2007}. 

In the following we derive the equations implemented, describe the computational scheme, give the expressions for magneto-optial properties measured experimentally and finally discuss the results of calculations for molecules and solids.

\section*{Results and Discussion}
\subsection*{One-particle density matrix in electromagnetic fields}  
Let us consider the response to uniform magnetic and electric fields. We use the temporal gauge, in which both of these fields are described by the vector potential $\mathbf{A}$ and are given by $\mathbf{B} = \nabla \times \mathbf{A}$  and $\mathbf{E} = -c^{-1} \partial_t \mathbf{A}$, respectively, where $c$ is the speed of light (atomic units are used throughout the paper). Though the fields are uniform, the vector potential $\mathbf{A}$ entering in the Hamiltonian $H$ is non-periodic. This gives rise to ill-defined expectation values of quantum mechanical operators describing physical properties of the system in the periodic basis. However, it turns out that for any operator $\mathcal{O}=\mathcal{O}_{\mathbf{r}_1\mathbf{r}_2}$ defined for two points $\mathbf {r}_1$ and $\mathbf {r}_2$ in real space it is possible to distinguish the periodic and gauge-invariant counterpart $\tilde{\mathcal{O}}=\tilde{\mathcal{O}}_{\mathbf{r}_1\mathbf{r}_2}$ by factoring out the Aharonov-Bohm-type phase\cite{chen2011,gonze2011,essin2010} 
\begin{equation}\label{eq0}
\varphi_{12}=-c^{-1}\int_{\mathbf {r}_2}^{\mathbf {r}_1}\mathbf{A}(\mathbf {r}) \mathrm{d} \mathbf {r}
\end{equation} so that 
\begin{equation}\label{eq1}
	\mathcal{O}_{\mathbf{r}_1\mathbf{r}_2} =  \tilde{\mathcal{O}}_{\mathbf{r}_1\mathbf{r}_2}\mathrm{exp}\left(i \varphi_{12} \right).
\end{equation}
Here we take $\hbar = e = 1$ and the integral is taken along the straight line between points $\mathbf {r}_2$ and $\mathbf {r}_1$ so that $\mathbf {r} = \mathbf {r}_2 + (\mathbf {r}_1 - \mathbf {r}_2) \xi$, $0 \le \xi \le 1$. 

This approach was previously used to derive corrections to the gauge-invariant counterpart $\tilde{\rho}$ of the one-particle density matrix $\rho$ in the static magnetic field~\cite{essin2010,gonze2011}. In the present paper we generalize these derivations to the case of time-dependent electromagnetic fields by rewriting the time-dependent Liouville equation 
\begin{equation} \label{eq2}
\begin{split}
- i \partial_t \rho + [H,\rho]=0
\end{split}
\end{equation}
 in terms of $\tilde{\rho}$. Here and below the commutator of two operators $\mathcal{O}^{(1)}$ and $\mathcal{O}^{(2)}$ is introduced as
\begin{equation} \label{eq3}
\begin{split}
[\mathcal{O}^{(1)},\mathcal{O}^{(2)}]&_{\mathbf{r}_1\mathbf{r}_3}=\\
&\int\mathrm{d} \mathbf{r}_2\left(\mathcal{O}^{(1)}_{\mathbf{r}_1\mathbf{r}_2}\mathcal{O}^{(2)}_{\mathbf{r}_2\mathbf{r}_3}-\mathcal{O}^{(2)}_{\mathbf{r}_1\mathbf{r}_2}\mathcal{O}^{(1)}_{\mathbf{r}_2\mathbf{r}_3}\right).
\end{split}
\end{equation}

Using Eq.~(\ref{eq1}) for the relation between $\tilde{\rho}$ and $\rho$ in real space, the time-dependent Liouville equation (\ref{eq2})  gives
\begin{equation} \label{eq4}
\begin{split}
& -i e^{i \varphi_{13}}\left( \partial_t  + i \partial_t  \varphi_{13} \right) \tilde{\rho}_{\mathbf{r}_1\mathbf{r}_3} = \\
& \displaystyle \int \mathrm{d}\mathbf{r}_2  e^{i (\varphi_{12}+\varphi_{23})}\left( \tilde{\rho}_{\mathbf{r}_1\mathbf{r}_2}  \tilde{H}_{\mathbf{r}_2\mathbf{r}_3} - \tilde{H}_{\mathbf{r}_1\mathbf{r}_2} \tilde{\rho}_{\mathbf{r}_2\mathbf{r}_3} \right).
\end{split}
\end{equation}
It should be noted that $\tilde{H}=H_0+\delta \tilde{H}$, where the difference $\delta \tilde{H}$ between the gauge-invariant counterpart $\tilde{H}$ of the Hamiltonian and unperturbed Hamiltonian $H_0$ is related to the local-field effects coming from changes in the electron density induced by the external fields and corresponds to the variation of Hartree and exchange-correlation potentials in TDDFT (see page 1 of Supplementary information). 

Eq. (\ref{eq4}) is equivalent to
\begin{equation} \label{eq5}
\begin{split}
& -i \left( \partial_t  + i \partial_t  \varphi_{13} \right) \tilde{\rho}_{\mathbf{r}_1\mathbf{r}_3} = \\
& \displaystyle \int \mathrm{d}\mathbf{r}_2  e^{i \varphi_{123}}\left( \tilde{\rho}_{\mathbf{r}_1\mathbf{r}_2}  \tilde{H}_{\mathbf{r}_2\mathbf{r}_3} - \tilde{H}_{\mathbf{r}_1\mathbf{r}_2} \tilde{\rho}_{\mathbf{r}_2\mathbf{r}_3} \right),
\end{split}
\end{equation}
\noindent where $\varphi_{123} = \varphi_{12} +  \varphi_{23} + \varphi_{31}$. 

This phase corresponds to the flux of the magnetic field through the triangle formed by points $\mathbf{r}_1$, $\mathbf{r}_2$ and $\mathbf{r}_3$: 
\begin{equation} \label{eq5_1}
\begin{split}
\varphi_{123} =\frac{1}{2c} \mathbf{B}\cdot  (\mathbf {r}_1 -\mathbf {r}_2)  \times (\mathbf {r}_2 -\mathbf {r}_3).
\end{split}
\end{equation}

The time derivative of the phase $\varphi_{13}$ on the left-hand side of Eq.~(\ref{eq5}) introduces the electric field 
\begin{equation} \label{eq6}
\begin{split}
\partial_t \varphi_{13}=\mathbf{E} \cdot (\mathbf{r}_1 -\mathbf{r}_3).
\end{split}
\end{equation}

Combining Eqs. (\ref{eq5})--(\ref{eq6}), we arrive at
\begin{equation} \label{eq7}
\begin{split}
-i ( \partial_t  + i  \mathbf{E} &\cdot (\mathbf{r}_1 -\mathbf{r}_3) ) \tilde{\rho}_{\mathbf{r}_1\mathbf{r}_3} = \\
 \displaystyle \int \mathrm{d}\mathbf{r}_2  &e^{i \mathbf{B}\cdot  (\mathbf {r}_1 -\mathbf {r}_2)  \times (\mathbf {r}_2 -\mathbf {r}_3) /2c}\\
&\cdot\left( \tilde{\rho}_{\mathbf{r}_1\mathbf{r}_2}  \tilde{H}_{\mathbf{r}_2\mathbf{r}_3} - \tilde{H}_{\mathbf{r}_1\mathbf{r}_2} \tilde{\rho}_{\mathbf{r}_2\mathbf{r}_3} \right).
\end{split}
\end{equation}
This expression is gauge-invariant and includes all corrections to time-dependent electric and magnetic fields. Therefore, it can be used to derive expressions for responses of any order to electromagnetic fields.

To describe magneto-optical effects on the basis of Eq. (\ref{eq7}) we assume that $\mathbf{E}$ corresponds to the oscillating electric field of the electromagnetic wave and $\mathbf{B}$ to the static magnetic field applied. The magnetic field of the electromagnetic wave is neglected. We, therefore, consider only the first-order corrections in $\mathbf{E}$, $\mathbf{B}$ and $\mathbf{E} \times \mathbf{B}$. Keeping only the terms to the first order in the magnetic field is reasonable even for strong magnetic fields $B \ll c/a^2 \sim10^5$ T, where $a = 1$ \AA~is taken as a typical interatomic distance. 

Eq. (\ref{eq7}) for the density matrix then takes the form
\begin{equation} \label{eq8}
\begin{split}
-i \partial_t  &\tilde{\rho}_{\mathbf{r}_1\mathbf{r}_3}- \int \mathrm{d}\mathbf{r}_2  \cdot\left( \tilde{\rho}_{\mathbf{r}_1\mathbf{r}_2}  \tilde{H}_{\mathbf{r}_2\mathbf{r}_3} - \tilde{H}_{\mathbf{r}_1\mathbf{r}_2} \tilde{\rho}_{\mathbf{r}_2\mathbf{r}_3} \right) \\ &=  -\mathbf{E} \cdot (\mathbf{r}_1 -\mathbf{r}_3) \tilde{\rho}_{\mathbf{r}_1\mathbf{r}_3}\\ &+ \frac{i}{2c} \int \mathrm{d}\mathbf{r}_2 \   \mathbf{B}\cdot  (\mathbf {r}_1 -\mathbf {r}_2)  \times (\mathbf {r}_2 -\mathbf {r}_3)  \\ &\quad \quad\quad\quad\quad\cdot\left( \tilde{\rho}_{\mathbf{r}_1\mathbf{r}_2}  \tilde{H}_{\mathbf{r}_2\mathbf{r}_3} - \tilde{H}_{\mathbf{r}_1\mathbf{r}_2} \tilde{\rho}_{\mathbf{r}_2\mathbf{r}_3} \right).
\end{split}
\end{equation}

Using that $\mathcal{O}_{\mathbf{r}_1\mathbf{r}_2}(\mathbf{r}_1-\mathbf{r}_2)=[\mathbf{r},\mathcal{O}]_{\mathbf{r}_1\mathbf{r}_2}$ and introducing notations for the anticommutator of two operators $\mathcal{O}^{(1)}$ and $\mathcal{O}^{(2)}$ 
\begin{equation} \label{eq9}
\begin{split}
\{\mathcal{O}^{(1)},\mathcal{O}^{(2)}\}&_{\mathbf{r}_1\mathbf{r}_3}=\\
&\int\mathrm{d} \mathbf{r}_2\left(\mathcal{O}^{(1)}_{\mathbf{r}_1\mathbf{r}_2}\mathcal{O}^{(2)}_{\mathbf{r}_2\mathbf{r}_3}+\mathcal{O}^{(2)}_{\mathbf{r}_1\mathbf{r}_2}\mathcal{O}^{(1)}_{\mathbf{r}_2\mathbf{r}_3}\right)
\end{split}
\end{equation}
and velocity operator $\mathbf{V}=-i[\mathbf{r},\tilde{H}]$ computed with account of all non-local contributions to the Hamiltonian, such as from non-local pseudopotentials, Eq. (\ref{eq8}) can be finally rewritten as
\begin{equation} \label{eq10}
\begin{split}
 -i \partial_t  \tilde{\rho} + [\tilde{H},\tilde{\rho}]= -\frac{1}{2} \left\{
\mathbf{E}+\frac{1}{c} \mathbf{V} \times \mathbf{B}, [\mathbf{r},\tilde{\rho}] \right\}.
\end{split}
\end{equation}
This is simply the quantum Bolzmann equation with the Lorentz driving force on the right-hand side. Unlike the singular position operator $\mathbf{r}$, the commutator $ [\mathbf{r},\tilde{\rho}]$ of the position operator with the periodic function $\tilde{\rho}$ is well defined here and can be substituted by the derivative with respect to the wave vector, $ i\partial_{\mathbf{k}} \rho_{\mathbf{k}}$, in reciprocal space \cite{essin2010,chen2011,gonze2011}.

Moving the term coming from the local-field effects to the right-hand side,
\begin{equation} \label{eq10_1}
\begin{split}
 -i \partial_t  \tilde{\rho} + &[H_0,\tilde{\rho}]= \\
&-\frac{1}{2} \left\{
\mathbf{E}+\frac{1}{c} \mathbf{V} \times \mathbf{B}, [\mathbf{r},\tilde{\rho}] \right\} -  [\delta \tilde{H},\tilde{\rho}],
\end{split}
\end{equation}
we get all terms dependent on the external fields on the right-hand side of the equation. Differentiating the Liouville equation~(\ref{eq10_1}), one can evaluate the derivatives of the density matrix $\tilde{\rho}^{(P)} = \partial \tilde{\rho}/\partial P$ with respect to perturbations $P$ of parameters of the Hamiltonian, such as the electric field $\mathbf{E}$ or magnetic field $\mathbf{B}$.

\subsection*{Numerical solution of Liouville equation} 
In the following we consider solution of the Liouville equation~(\ref{eq10_1}) within TDDFT, i.e. assuming that $\rho$ is the Kohn-Sham density matrix and $H$ is the Kohn-Sham Hamiltonian. The same Liouville equation, however, describes magneto-optical effects in any other first-principles framework and a similar computational scheme can be used.

From the computational point of view, it is convenient to divide the $n$-th order derivative $\tilde{\rho}^{(P)}$ of the density matrix describing the joint response to the perturbations $P=P_1P_2...P_n$ into four  blocks within and between the occupied  (V)  and unoccupied subspaces (C):
 \begin{equation} \label{eq11}
\begin{split}
\tilde{\rho}^{(P)}=
\tilde{\rho}^{(P)}_{\mathrm{VV}}+
\tilde{\rho}^{(P)}_{\mathrm{CC}}+
\tilde{\rho}^{(P)}_{\mathrm{VC}}+
\tilde{\rho}^{(P)}_{\mathrm{CV}}.
\end{split}
\end{equation}
These blocks correspond to $\tilde{\rho}^{(P)}_{\mathrm{VV}}=P_v\tilde{\rho}^{(P)}P_v$, $\tilde{\rho}^{(P)}_{\mathrm{CC}}=P_c\tilde{\rho}^{(P)}P_c$,
$\tilde{\rho}^{(P)}_{\mathrm{VC}}=P_v\tilde{\rho}^{(P)}P_c$,
and $\tilde{\rho}^{(P)}_{\mathrm{CV}}=P_c\tilde{\rho}^{(P)}P_v$, where $P_{v} =\rho^{(0)}$ and $P_{c} = 1 - P_{v} $ are the projectors onto the occupied and unoccupied bands.

Following the density matrix perturbation theory~\cite{lazzeri2003}, to get the elements  of the derivative of the density matrix $\tilde{\rho}^{(P)}_{\mathrm{CV}}$ between the unoccupied and occupied subspaces, we project the Liouville equation~(\ref{eq10_1}) onto unperturbed Kohn-Sham wavefunctions $ | \psi_{v\mathbf{k}}^{(0)} \rangle$ of occupied bands $v$:
\begin{equation} \label{eq12}
\begin{split}
L_{v\mathbf{k}} (\Omega)  | \eta_{v\mathbf{k}}^{(P)} \rangle &= P_c R^{(P)} [\tilde \rho^{(n-1)},...,\rho^{(0)},n^{(P)}]  | \psi_{v\mathbf{k}}^{(0)} \rangle.
\end{split}
\end{equation}
Here the operator on the left-hand side is given by $L_{v\mathbf{k}} (\Omega)  = \Omega + H_0 - \epsilon_{v \mathbf{k}}$, where $\Omega$ is frequency considered and $\epsilon_{v \mathbf{k}}$ is the energy of the unperturbed state $ | \psi_{v\mathbf{k}}^{(0)} \rangle$. The operator $R$ on the right-hand side includes all terms dependent on the perturbation $P$ coming from the right-hand side of Eq.~(\ref{eq10_1})  and is determined by the derivatives of the density matrix of the previous orders (see equations for each type of perturbation on pages 1--3 of Supplementary Information). If the local-field effects are taken into account, it also depends on the derivative of the electron density $n^{(P)}$ to the perturbation $P$, $n^{(P)}(\mathbf {r}_1)= {\rho}^{(P)}(\mathbf {r}_1,\mathbf {r}_2) \delta (\mathbf {r}_1-\mathbf {r}_2)$  (see page 1 of Supplementary information).

The solution of Eq. (\ref{eq12}) corresponds to
\begin{equation} \label{eq13}
\begin{split}
| \eta_{v\mathbf{k}}^{(P)}(\Omega) \rangle= P_{c}  \tilde{\rho}^{(P)} ( \Omega) | \psi_{v\mathbf{k}}^{(0)} \rangle = \tilde{\rho}^{(P)}_{\mathrm{CV}} ( \Omega) | \psi_{v\mathbf{k}}^{(0)} \rangle
\end{split}
\end{equation}
and once it is known, the elements $\tilde{\rho}^{(P)}_{\mathrm{CV}}$ of the derivative of the density matrix  between unoccupied and occupied subspaces can be computed as 
\begin{equation} \label{eq14}
\begin{split}
\tilde{\rho}^{(P)}_{\mathrm{CV}} ( \Omega) = \int_\mathrm{BZ}  \frac{\mathrm{d}\mathbf{k}}{(2\pi)^{3}} \sum_{v} | \eta^{(P)}_{v\mathbf{k}} ( \Omega)  \rangle \langle \psi_{v\mathbf{k}}^{(0)} |.
\end{split}
\end{equation}
The elements between the occupied and unoccupied subspaces can be found as $\tilde{\rho}^{(P)}_{\mathrm{VC}}  ( \Omega) = (\tilde{\rho}^{(P)}_{\mathrm{CV}} (-\Omega^*))^*$ and to obtain them, Eq. (\ref{eq12}) should be also solved for the frequency $-\Omega^*$. If the local-field effects are taken into account, Eq. (\ref{eq12})  has to be solved self-consistently as the derivative $\tilde{\rho}^{(P)}$ of the density matrix  determines the derivative of the electron density $n^{(P)}$, which enters on the right-hand side of Eq. (\ref{eq12}).

Solution of Eq. (\ref{eq12}) is performed in the present paper using the efficient Sternheimer approach~\cite{andrade2007,strubbe_tesis,strubbe2012,andrade2015}, which corresponds to the iterative search of the function $| \eta_{v\mathbf{k}}^{(P)}(\Omega) \rangle$ that fits into this equation at each frequency $\Omega$. Other approaches, such as sum over states \cite{seth2008}, methods based on Casida's equation \cite{seth2008, seth2008a}, complex polarization propagator \cite{solheim2008a, solheim2008b} and real-time propagation \cite{lee2011} have been used to compute absorption and magneto-optical spectra of molecules. The sum over states, Casida's equation \cite{casida} and complex polarization propagator \cite{norman2001, norman2005}, however, require inclusion of many well converged unoccupied states. Such calculations are not feasible for large systems, where too many KS states should be computed. They also fail to describe properly high-energy excitations due to poor convergence of the corresponding KS states. Casida's equation \cite{casida} furthermore relies on the use of real wavefunctions and cannot be straightforwardly extended to solids, where KS states are complex.

Neither Sternheimer approach \cite{andrade2007,strubbe_tesis,strubbe2012,andrade2015}, nor real-time propagation \cite{lee2011} need calculation of unoccupied states. They also have a favourable scaling of $O(N^2)$ with the system size $N$ as compared,  for example, to $O(N^3)$ for the sum over states (Refs. \onlinecite{strubbe_tesis,andrade2007,lee2011}). The advantage of the real-time propagation is that it makes possible calculation of responses for all frequencies at once. However, long propagation times are required to achieve a good resolution. The Sternheimer approach is more appropriate for computing  the spectra in a narrow frequency region with a high resolution. The calculations for different frequencies can be performed in parallel. Most importantly, it is ideally suited for implementation of the density matrix perturbation theory considered in the present paper (see Eq. (\ref{eq12})).

A small but finite imaginary frequency $\delta$ is added to the frequency $\Omega_0$  of the external perturbation to avoid divergences at resonances~\cite{andrade2007,strubbe_tesis,strubbe2012,andrade2015,norman2001, norman2005} so that  $\Omega = \Omega_0 + i\delta$. This imaginary frequency $\delta$ determines the linewidth in the calculated spectra.

To find the derivatives to the density matrix within the occupied, $\tilde{\rho}^{(P)}_ {\mathrm{VV}}$, and unoccupied, $ \tilde{\rho}^{(P)}_{\mathrm{CC}}$, subspaces, one can, in principle, also look for solution of the Liouville equation~(\ref{eq10}). However, in the case when the density matrix is idempotent, like the Kohn-Sham density matrix, the solution can be found explicitly from the idempotency condition, $\rho = \rho \rho $, and this reduces considerably the computational cost. The idempotency condition in terms of the periodic counterpart $\tilde{\rho}$ of the density matrix and to the first order in the magnetic field can be written as~\cite{essin2010,gonze2011} 
\begin{equation} \label{eq15}
\begin{split}
&\tilde{\rho}=\tilde{\rho}\tilde{\rho}+
 \frac{i}{2c} \mathbf{B} \cdot [\mathbf{r},\tilde{\rho}] \times [\mathbf{r},\tilde{\rho}].
\end{split}
\end{equation}
The commutator $[\mathbf{r},\tilde{\rho}]$ corresponding to $ i\partial_{\mathbf{k}} \tilde{\rho}_{\mathbf{k}}$ in reciprocal space is determined in the present paper within the $\mathbf{k} \cdot \mathbf{p}$  theory~\cite{strubbe_tesis,strubbe2012,andrade2015} (see equations on pages 2 and 3 Supplementary information). 

The polarizability $\alpha_{0 \nu \mu}$ in the absence of the magnetic field and the contribution $\alpha_{\nu \mu, \gamma}$  to the polarizability in the presence of the magnetic field ($\alpha_{\nu \mu}=\alpha_{0 \nu \mu}+\alpha_{\nu \mu, \gamma} B_\gamma$) are obtained from the current response as
\begin{equation} \label{eq16}
\begin{split}
\alpha_{0\nu \mu} (\Omega)  = \frac{i}{\Omega} \mathrm{Tr} \left[V_{\nu} \tilde\rho^{(E_\mu)} (\Omega) \right]
\end{split}
\end{equation}
and
\begin{equation} \label{eq17}
\begin{split}
\alpha_{\nu \mu, \gamma} (\Omega)  = \frac{i}{\Omega} \mathrm{Tr} \left[V_{\nu} \tilde\rho^{(E_\mu B_\gamma)} (\Omega) \right].
\end{split}
\end{equation}
These polarizabilies can be used to compute the experimentally measurable physical properties as described below.

\subsection*{Experimentally measured properties} 
The capacity of the system to absorb light is characterized using absorbance $A=-\log (I/I_0)$, which is defined through the ratio of intensities of the incident, $I_0$, and transmitted light, $I$. The magnitudes of the electric field vectors in the transmitted, $E$, and incident light, $E_0$, at frequency $\Omega_0$ are related as $E=E_0 \exp(-n'\Omega_0l/c)$, where $n'$ is the imaginary part of the refractive index $n'=\mathrm{Im} \  n$ and $l$ is the distance passed by the light through the sample studied. Since $I\sim E^2$, it can be stated that
\begin{equation} \label{eq18}
\begin{split}
 A=\frac{2n'\Omega_0l}{c\ln10}.
\end{split}
\end{equation}

The difference in the absorbance of the left ($+$) and right ($-$) circularly polarized light corresponds to the MCD response and is determined by the difference in the refractive indices $n'_+-n'_-$ for these two light components:
\begin{equation} \label{eq19}
\begin{split}
\Delta A=A_+-A_-=\frac{2(n'_+-n'_-)\Omega_0l}{c\ln10}.
\end{split}
\end{equation}

The refractive index $n$ is determined by the equation 
\begin{equation} \label{eq20_00}
\begin{split}
\epsilon_{\nu\mu}E_\mu=n^2E_\mu,
\end{split}
\end{equation} 
where $\epsilon_{\nu\mu}$ is the dielectric tensor. For crystals, the dielectric tensor is related to the electric susceptibility $\chi_{\nu \mu}$ as
\begin{equation} \label{eq20_0}
\begin{split}
\epsilon_{\nu \mu}=\delta_{\nu \mu}+4\pi\chi_{\nu \mu}. 
\end{split}
\end{equation}
The latter corresponds to the polarizability per unit volume so that $\chi_{\nu \mu}=\alpha_{\nu \mu}/w$, where $w$ is the unit cell volume and $\alpha_{\nu \mu}$ is given by Eqs. (\ref{eq16}) and (\ref{eq17}).

In the case when the light propagation takes place along the optical axis $z$ and no birefingence is observed, the refractive index in the absence of the magnetic field  is equal to $n_0=\epsilon_{xx}^{1/2}=\epsilon_{yy}^{1/2}$. The magnetic field provides just a small correction to this refractive index and it can be shown from Eqs. (\ref{eq20_00}) and (\ref{eq20_0})  (see pages 5 and 6 of Supplementary information) that
\begin{equation} \label{eq20}
\begin{split}
n^{\pm}-n_0 \approx \pm i\frac{2\pi \chi_{xy}}{n_0}.
\end{split}
\end{equation}

Using Eq. \ref{eq19}, the difference in the absorbance of the left and right circularly polarized light can be found as 
\begin{equation} \label{eq21}
\begin{split}
\Delta A_z=\frac{4\pi  \Omega_0 l}{c\ln10}  \mathrm{Re}  \left[ \frac{ \chi_{xy} (\Omega)-\chi_{yx} (\Omega)}{n_0} \right].
\end{split}
\end{equation}

Note that ellipticity $\theta=(E_+-E_-)/(E_++E_-)$ gained by the linearly polarized light is  different just by a numerical coefficient $\theta=\Delta A_z (\ln10)/4$. The angle of Faraday rotation is determined by a similar expression as $\theta$ but with the imaginary part of $\chi_{\nu \mu}$ instead of the real one \cite{barron2004} (see page 5 of Supplementary information). By contrast, in the magneto-optical polar Kerr effect for reflected light, the ellipticity and angle of rotation are determined by  $\mathrm{Im} \ \chi_{\nu \mu}$ and  $\mathrm{Re} \  \chi_{\nu \mu}$, respectively \cite{agranovich1991}.

For molecules, the measurements are usually performed for a small concentration of randomly oriented molecules immersed into a transparent solvent or in vacuum. In this case, the total dielectric tensor of the medium can be presented as 
\begin{equation} \label{eq22_1}
\begin{split}
\epsilon_{\nu \mu}=n_S^2\delta_{\nu \mu}+4\pi\bar\alpha_{\nu \mu}N, 
\end{split}
\end{equation}
where $\delta_{\nu \mu}$ is the Kronecker delta, $n_S$ is the refractive index of the solvent or vacuum, $\bar\alpha_{\nu \mu}$ is the orientationally averaged polarizabiltiy of the molecules and $N$ is their number density. The orientationally averaged polarizability is given by 
\begin{equation} \label{eq22}
\begin{split}
\bar\alpha_{\nu \mu}=\frac{1}{3}\alpha_{0aa}\delta_{\nu \mu}+\frac{1}{6}B e_{abc}\alpha_{ab,c}e_{\nu \mu},
\end{split}
\end{equation}
 where $e_{\nu\mu}$ and $e_{abc}$  are the Levi-Civita tensors of the second and third order, respectively, and the polarizabilities $\alpha_{0aa}$ and $\alpha_{ab,c}$ are computed from Eqs. (\ref{eq16}) and (\ref{eq17}) considering internal molecular axes. 

For molecules, it is common to use molar extinction coefficients $\epsilon=A/Cl$, i.e. absorbance per unit length and molar concentration. The molar concentration $C$ in this expression is related to the number density as $C=N/N_\mathrm{A}$, where $N_\mathrm{A}$ is the Avogadro constant. Taking into account that the concentration of the molecules is small, the refractive index in the absence of the magnetic field becomes approximately $n_0 \approx n_S+2\pi N\alpha_{0aa}/(3n_S)$ and this gives the molar extinction coefficient
\begin{equation} \label{eq23}
\begin{split}
 \epsilon=\frac{4\pi\Omega_0 N_\mathrm{A} }{3n_Sc\ln10}\mathrm{Im} \  \alpha_{0aa}.
\end{split}
\end{equation}

The refractive indices for the left and right circularly polarized light can be correspondingly expressed as 
\begin{equation} \label{eq24}
\begin{split}
n^{\pm}-n_0 \approx \pm i\frac{\pi N B }{3n_S}e_{abc}\alpha_{ab,c}.
\end{split}
\end{equation}
The difference $\Delta\epsilon$ in the molar extinction coefficients for the left and right circularly polarized light per unit magnetic field can, therefore, be found as
\begin{equation} \label{eq25}
\begin{split}
\Delta\epsilon=\frac{\Delta A}{BCl}=  \frac{4\pi N_\mathrm{A} }{3n_Sc\ln10}e_{abc}\mathrm{Re} \  \alpha_{ab,c}.
\end{split}
\end{equation}

The formalism for calculation of the magneto-optical response proposed in the present paper and expressions for the physical properties listed above have been implemented in the Octopus code \cite{marques2003,castro2006,andrade2015}. The results of the  tests for molecules and solids are presented below.

\begin{figure*}
	\centering
	\includegraphics[width=0.9\textwidth]{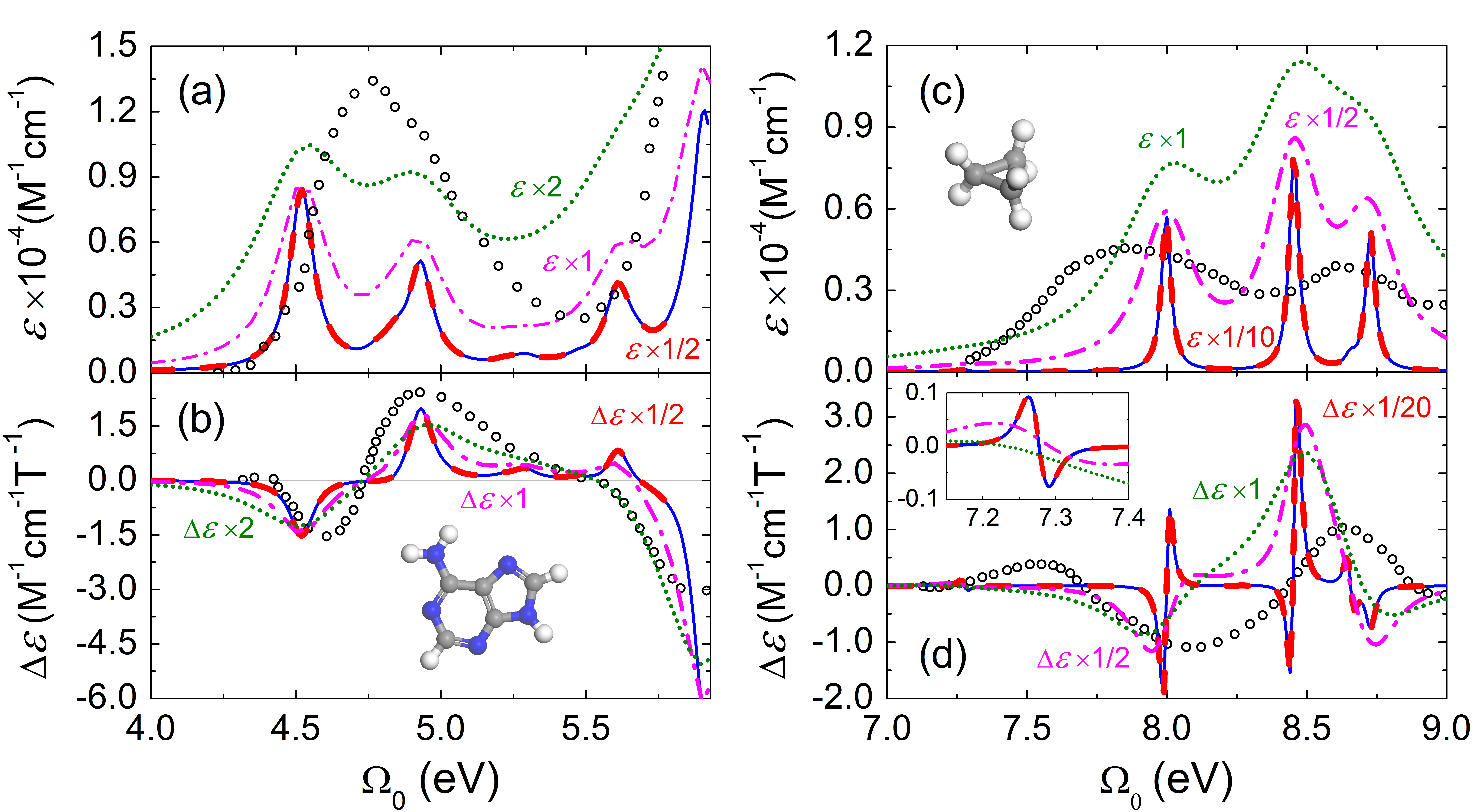}
	\caption{Molar extinction coefficient $\epsilon$ (\textbf{a} and \textbf{c}, in M$^{-1}$cm$^{-1}$) and difference $\Delta\epsilon$ in the molar extinction coefficients for the left and right circularly polarized light per unit magnetic field  (\textbf{b} and \textbf{d}, in M$^{-1}$cm$^{-1}$T$^{-1}$) for adenine (\textbf{a} and \textbf{b}, $\delta = 0.05$ eV) and cyclopropane (\textbf{c} and \textbf{d}, $\delta = 0.02$ eV) as functions of the frequency of light $\Omega_0$ (in eV) calculated using the present solid-state formalism (solid blue lines) and standard finite-system formulation (red dashed lines). The corresponding curves are virtually indistinguishable. The results obtained in the finite-system formulation for linewidths $\delta = 0.1$ eV and $\delta = 0.2$ eV are shown by magenta dash-dotted lines and green dotted lines, respectively. The experimental data for adenine \cite{sutherland1984} in water and cyclopropane \cite{gedanken1976} in the gas phase are represented by circles. To show the results for different linewidths and experimental data on the same scale, the following scaling factors are introduced: 2, 1, 1/2 for the linewidths of 0.2 eV, 0.1 eV and 0.05 eV for the absorption and MCD spectra of adenine, 1, 1/2, 1/10 for the linewidths of 0.2 eV, 0.1 eV and 0.02 eV for the absorption spectra of cyclopropane and 1, 1/2, 1/20, respectively, for the MCD spectra of cyclopropane. In the calculations for adenine, the refractive index of water is taken equal to 1.35 (Ref. \onlinecite{hale1973}).
 The parts of the spectra shown lie below the ionization potential at zero temperature (6.7 eV and 9.4 eV for adenine and cyclopropane, respectively, according to our calculations). Carbon, hydrogen and nitrogen atoms in the atomistic structures are coloured in gray, white and blue, respectively. The inset shows the first MCD peak of cyclopropane. }
	\label{fig:abs}
\end{figure*}

\subsection*{Results of calculations for molecules} 
First the tests of the developed formalism were performed for molecules (Fig. \ref{fig:abs}) in a large simulation box with periodic boundary conditions. Traditionally the MCD response of molecules is divided into $\mathcal{A}$ and  $\mathcal{B}$ terms  (see equations on pages 4 and 5 of Supplementary information). The $\mathcal{B}$ term \cite{solheim2008a, lee2011, seth2008} comes from perturbations of molecular states in the magnetic field and is present in all systems. The $\mathcal{A}$ term \cite{solheim2008a, lee2011, seth2008a} comes from perturbations of energies of excited states with non-zero orbital angular momenta. Such states are present only in molecules with rotational symmetry at least of the third order. Since transitions to states with opposite orbital angular momenta are coupled to the light of different polarization, Zeeman splitting leads to an energy shift between absorption peaks for the left and right circularly polarized light. The MCD response in this case is described by the derivative of the spectral density \cite{seth2008, seth2008a} and has second-order poles.

To check that both $\mathcal{A}$ and $\mathcal{B}$ terms are well described within the developed formalism, we have performed the calculations for adenine and cyclopropane (Fig. \ref{fig:abs}). Adenine is not symmetric and only the $\mathcal{B}$ term contributes to the magneto-optical response. Though we use the simplest local-density approximation (LDA) \cite{perdew1981} for the exchange-correlation contribution to the electron energy and adiabatic approximation (ALDA) for the response, we find that the changes in the sign of the MCD signal for adenine are properly described as compared to the experimental data \cite{sutherland1984} (Fig. \ref{fig:abs}b). The magnitudes of the peaks for the simple optical absorption and the $\mathcal{B}$ term of the magneto-optical response scale inversely proportional to the linewidth, which is an input parameter of our calculations. Using a reasonable linewidth of $\delta = 0.1$ eV, we get the absorption (Fig. \ref{fig:abs}a) and MCD (Fig. \ref{fig:abs}b) spectra with the magnitude of the peaks comparable to the experimental ones. 

Cyclopropane has a rotational symmetry of the third order and its magneto-optical response has both $\mathcal{A}$ and $\mathcal{B}$ contributions. We find that the $\mathcal{A}$ term is clearly dominant for cyclopropane at linewidth $\delta = 0.02$ eV (Fig. \ref{fig:abs}d), in agreement with previous calculations \cite{solheim2008a}. However, the $\mathcal{A}$ and $\mathcal{B}$ terms scale differently with the linewidth. $\mathcal{B}$ term is inversely proportional to the linewidth, while the $\mathcal{A}$ is inversely proportional to square of the linewidth. Therefore, raising the linewidth to the experimental values of $\delta = 0.1-0.2$ eV decreases the $\mathcal{A}$ term relative to the $\mathcal{B}$ term. For these linewidths, the shapes of the calculated curves and the magnitudes of the peaks approach the experimental ones \cite{gedanken1976}  (Figs. \ref{fig:abs}c and d). 

The calculations for the molecules (Fig. \ref{fig:abs}) demonstrate that the present formalism gives the results indistinguishable from the formulation using the position operator $\mathbf{r}$ (see page 4 of Supplementary information), which is commonly applied in literature for finite systems \cite{solheim2008a, solheim2008b, lee2011, seth2008, seth2008a}.

\subsection*{Results of calculations for solids}
To test the developed formalism for solids we have applied it to bulk silicon and a monolayer of hexagonal boron nitride. For these periodic systems, we set the linewidth at $\delta = 0.1$ eV, which is sufficient to resolve the important features of the spectra. Since we use LDA for our test calculations, the excitation energies are systematically underestimated. To adjust the position of the peaks we apply the scissor operator, i.e. rigidly shift the spectra, to include the correction to the band gap known from GW  calculations \cite{albrecht1998, botti2004, witz2005}. It should be, nevertheless, emphasized that the same code can be used with more advanced functionals like hybrid ones, which provide an improved description of the excitation energies. The approach can be also straightforwardly translated into the many-body framework. 

\begin{figure*}
	\centering
	\includegraphics[width=0.9\textwidth]{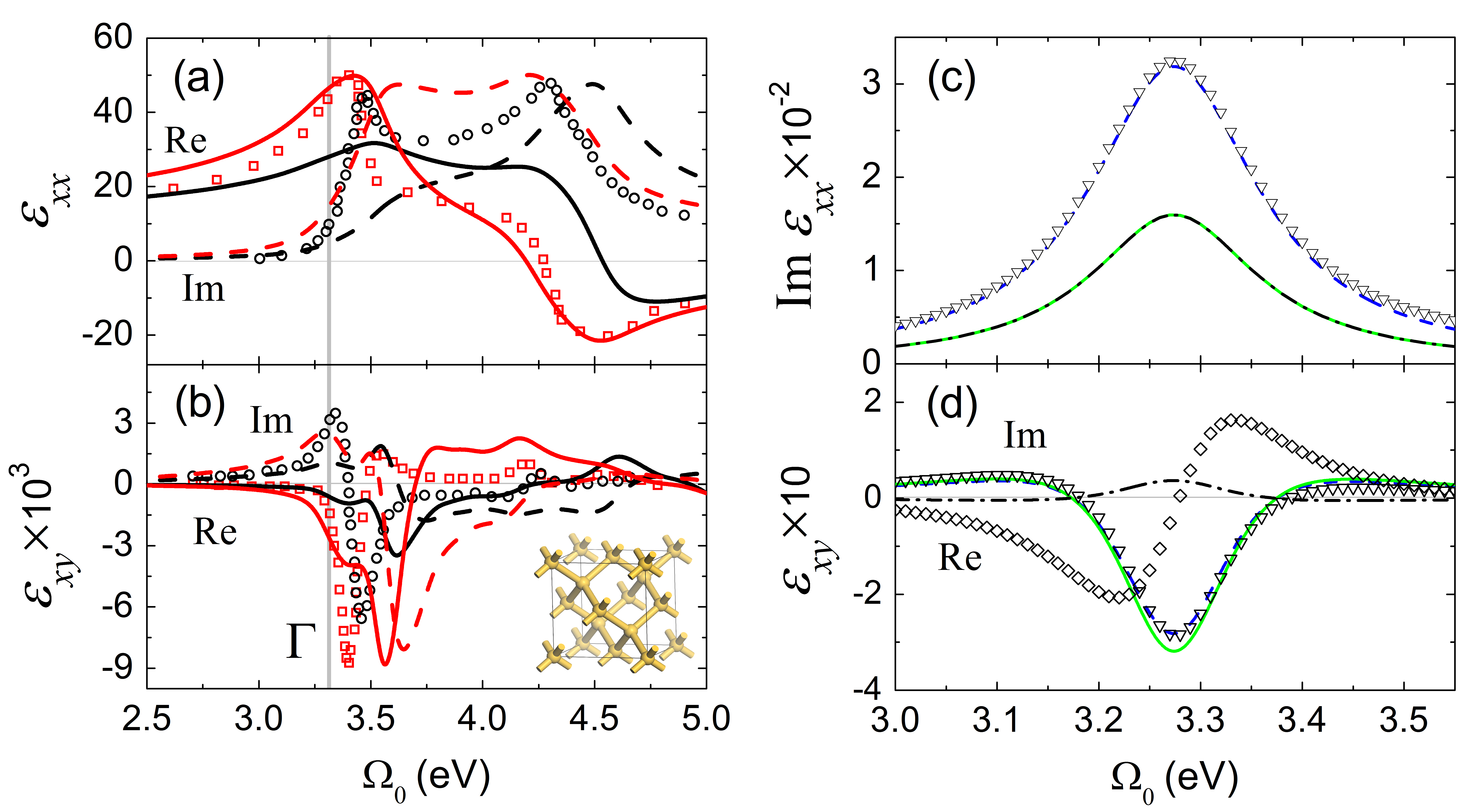}
	\caption{Calculated components $\epsilon_{xx}$ \textbf{a} and $\epsilon_{xy}$ \textbf{b} of the dielectric tensor of silicon as functions of  the frequency of  light $\Omega_0$ (in eV) for the magnetic field of 1 T along the $z$ axis. The real and imaginary parts are shown by solid and dashed lines, respectively. The results obtained with and without account of excitonic effects correspond to red and black lines, respectively. The calculated data are blue-shifted in energy by 0.7 eV to take into account the GW correction to the band gap \cite{albrecht1998, botti2004}. The experimental data from Refs. \onlinecite{lautenschlager1987} and \onlinecite{agranovich1991} for $\epsilon_{xx}$ and $\epsilon_{xy}$, respectively, are shown by symbols. The experimental data for $\epsilon_{xy}$ are scaled by a factor of 1/2. Squares correspond to the real parts and circles to the imaginary ones. The transitions at the $\Gamma$ point of the Brillouin zone are indicated by the vertical gray line. Calculated contributions to $\mathrm{Im} \  \epsilon_{xx}$ \textbf{c} and $\mathrm{Im} \  \epsilon_{xy}$ \textbf{d}  from the $\Gamma$ point: total contribution (triangles), contribution from all transitions  $\Gamma'_{25} \to \Gamma_{15}$ to the $\mathcal{A}$ term  (blue dashed lines) and contributions from transitions  $\Gamma'_{25} \to \Gamma_{15}$ with the magnetic quantum number $l_z=0 \to \pm 1$ (green solid lines) and  $ \pm 1 \to 0$  (black dash-dotted lines) to the $\mathcal{A}$ term. Total $\mathrm{Re} \  \epsilon_{xy}$ is shown by diamonds.}
	\label{fig:si}
\end{figure*}

While account of local-field effects through Eq. (\ref{eq10_1}) even within the simplest ALDA approximation is very important for molecules, for silicon and boron nitride, such adiabatic effects provide a minor correction to the spectra (see Fig. 2 of Supplementary information). The account of long-range exchange and correlation interactions in solids is, on the other hand, crucial for description of excitons. To take them into account we follow the approach proposed in Ref. \onlinecite{berger2015} in the TDCDFT framework. In this approach, non-adiabatic local-field effects are introduced through 
the exchange-correlation electric field 
\begin{equation}\label{eq2d}
\begin{split}
\mathbf{E}^\mathrm{xc}_\mathrm{mac}(\Omega)=\frac{i\Omega}{ w}\int_w \mathrm d\mathbf{r}  \int \mathrm d\mathbf{r}' \hat{f}_\mathrm{xc}(\mathbf{r},\mathbf{r}',\Omega)\delta\mathbf{j}(\mathbf{r}',\Omega),
\end{split}
\end{equation}
where tensor $\hat{f}_\mathrm{xc}(\mathbf{r},\mathbf{r}',\Omega)$ is  the TDCDFT exchange-correlation kernel and $\delta\mathbf{j}(\mathbf{r}',\Omega)$ is the induced current density. This field together with the macroscopic electric field $\mathbf{E}_\mathrm{mac}$ gives the macroscopic  Kohn-Sham electric field $\mathbf{E}^\mathrm{KS}_\mathrm{mac}=\mathbf{E}_\mathrm{mac}+\mathbf{E}^\mathrm{xc}_\mathrm{mac}$.

The macroscopic polarization 
\begin{equation}\label{eq2e}
\begin{split}
\mathbf{P}_\mathrm{mac}(\Omega)=\frac{-i}{\Omega w}\int_w \mathrm d\mathbf{r} \delta\mathbf{j}(\vec{r},\Omega),
\end{split}
\end{equation}
is related to the macroscopic Kohn-Sham electric field $\mathbf{E}^\mathrm{KS}_\mathrm{mac}$ through the Kohn-Sham electric susceptibility tensor $\hat{\chi}^\mathrm{KS}$ and to the macroscopic electric field $\mathbf{E}_\mathrm{mac}$ through the net susceptibility tensor $\hat{\chi}$:
\begin{equation}\label{eq2f}
\begin{split}
\mathbf{P}_\mathrm{mac}(\Omega)&=\hat{\chi}^\mathrm{KS}(\Omega)(\mathbf{E}_\mathrm{mac}(\Omega)+\mathbf{E}^\mathrm{xc}_\mathrm{mac}(\Omega))\\&=\hat{\chi}(\Omega)\mathbf{E}_\mathrm{mac}(\Omega)
\end{split}
\end{equation}

Neglecting microscopic current components in  Eq. (\ref{eq2d}), i.e. replacing the induced current  density $\delta\mathbf{j}(\vec{r}',\Omega)$ by its unit cell average, and using Eq. (\ref{eq2e}), the exchange-correlation electric field is written as
\begin{equation}\label{eq2g}
\begin{split}
\mathbf{E}^\mathrm{xc}_\mathrm{mac}(\Omega)=\hat{\beta}(\Omega)\mathbf{P}_\mathrm{mac}(\Omega),
\end{split}
\end{equation}
where 
\begin{equation}\label{eq2h}
\begin{split}
\hat{\beta}(\Omega)=-\frac{\Omega^2}{w}\int_w \mathrm  \mathrm d\mathbf{r}  \int \mathrm d\mathbf{r}' \hat{f}_\mathrm{xc}(\mathbf{r},\mathbf{r}',\Omega).
\end{split}
\end{equation}

Substitution of Eq. (\ref{eq2g}) into Eq. (\ref{eq2f}) gives
\begin{equation}\label{eq26}
\begin{split}
\frac{1}{\hat{\chi}(\Omega)}=\frac{1}{\hat{\chi}^\mathrm{KS}(\Omega)}-\hat{\beta}(\Omega).
\end{split}
\end{equation}

In the simplest case, $\hat{\beta}$ can be assumed static and isotropic, i.e. $\beta_{\nu \mu}=\beta\delta_{\nu \mu}$. Then the longitudial and transverse components of the electric susceptibility tensor are given by
\begin{equation}\label{eq27}
\begin{split}
\chi_{xx}(\Omega) =\frac{\chi^\mathrm{KS}_{xx}}{1-\beta\chi^\mathrm{KS}_{xx}(\Omega)}
\end{split}
\end{equation}
and 
\begin{equation}\label{eq28}
\begin{split}
\chi_{xy}(\Omega) \approx \frac{\chi^\mathrm{KS}_{xy}}{(1-\beta\chi^\mathrm{KS}_{xx}(\Omega))(1-\beta\chi^\mathrm{KS}_{yy}(\Omega))},
\end{split}
\end{equation}
respectively. In these expressions, we neglect the terms of the second order in the transverse components of $\hat{\chi}^\mathrm{KS}$. 

It should be noted that Eq. (\ref{eq27}) for the longitudinal response is equivalent to the head term of the long-range contribution (LRC) to the exchange-correlation kernel \cite{albrecht1998,botti2004,stubner2004} in TDDFT, which corresponds to $f_{xc}^{\mathrm{(LRC)}}(\mathbf{q}) = -\beta/q^2$ in reciprocal space. However, the latter model does not describe properly the transverse response. Eq. (\ref{eq28}) gives an adequate expression for the transverse response thanks to the tensorial nature of the exchange-correlation kernel $\hat{f}_\mathrm{xc}(\mathbf{r},\mathbf{r}',\Omega)$ in the TDCDFT framework.

\begin{figure*}
	\centering
	\includegraphics[width=0.9\textwidth]{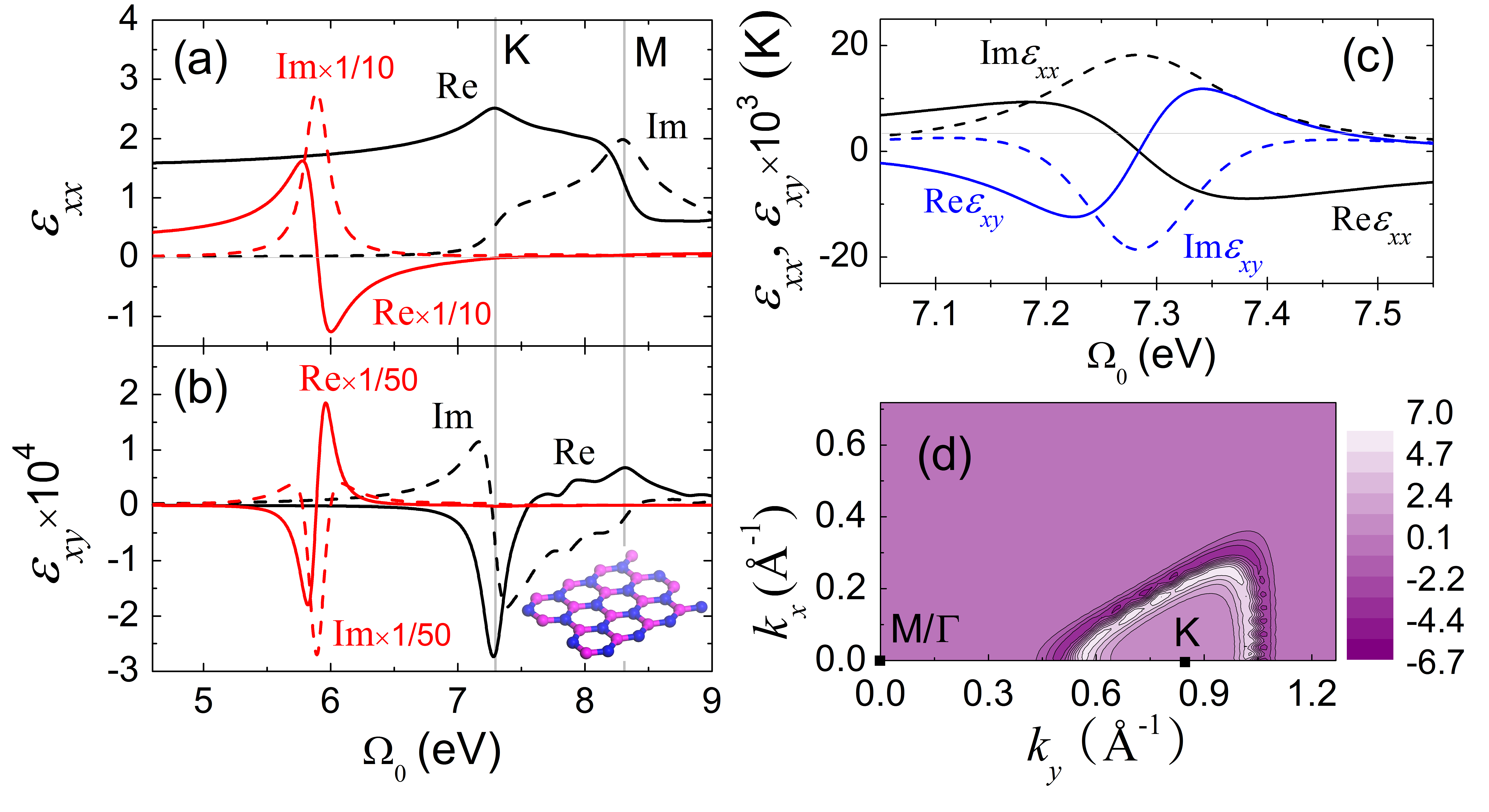}
	\caption{Calculated components $\epsilon_{xx}$ \textbf{a} and $\epsilon_{xy}$ \textbf{b} of the dielectric tensor of boron nitride monolayer as functions of  the frequency of light $\Omega_0$ (in eV) for the magnetic field of 1 T along the $z$ axis directed out of the plane. The real and imaginary parts are represented by solid and dashed lines, respectively. The results obtained with and without account of excitonic effects correspond to red and black lines, respectively. The data for $\epsilon_{xx}$ and $\epsilon_{xy}$ obtained with account of excitonic effects are multiplied by 1/10 and 1/50, respectively, to show all the results on the same scale. The calculated data are blue-shifted in energy by 2.6 eV to take into account the GW correction to the band gap \cite{witz2005}. The transitions at the K and M points of the Brillouin zone are indicated by vertical gray lines. Boron and nitrogen atoms in the atomistic structure are coloured in magenta and blue, respectively. \textbf{c} Calculated contributions to $\epsilon_{xx}$ (black lines) and $\epsilon_{xy}$ (blue lines) from the K points of the Brillouin zone. \textbf{d} Calculated contributions to ${\rm Re}\ \epsilon_{xy} \times 10^3$ from different points ($k_x$,$k_y$,0) (in \AA$^{-1}$) of the Brillouin zone of the 4-atom cell for  $\Omega_0=7.8$ eV. }
	\label{fig:bn}
\end{figure*}

Let us first discuss the results for bulk silicon (Fig. \ref{fig:si}). Fig. \ref{fig:si}b shows that the spectra $\mathrm{Re}/\mathrm{Im}\ \epsilon_{xy}$ for the transverse component of the dielectric tensor calculated even without account of excitonic effects follow qualitatively the shapes of the experimental curves \cite{agranovich1991} at the direct absorption edge.  The analysis of optical transitions at the $\Gamma$ point of the Brillouin zone, where the highest valence and lowest conduction  bands are formed by triply degenerate $p$-like states ($\Gamma'_{25}$ and $\Gamma_{15}$, respectively) \cite{cardona1966}, reveals significant contributions that can be attributed to the $\mathcal{A}$ term (Fig. \ref{fig:si}d). Two inequivalent contributions come from excitations with the change in the magnetic quantum number $l_z$ from 0 to $\pm 1$ and vice versa. The ratio $\epsilon_{xy}/\epsilon_{xx}$ for each of them at the resonance frequency characterizes the relative frequency shift in the magnetic field
\begin{equation}\label{eq29}
\begin{split}
 \frac{\epsilon_{xy}}{\epsilon_{xx}}  \sim \frac{\Delta m_z B_z \Delta l_z}{\delta}, 
\end{split}
\end{equation}
where $\Delta m_z$ is the change in the orbital magnetic dipole moment and $\Delta l_z$ is the change of the magnetic quantum number (see explanation on page 5 of Supplementary information). Correspondingly, we can estimate the effective g-factors $g=-\Delta m_z/\mu_\mathrm{B}\Delta l_z$, where $\mu_\mathrm{B}$ is the Bohr magneton, and they are found to be $g=3.5$ in $\Gamma'_{25} \to \Gamma_{15}$  transitons with $l_z = 0 \to \pm 1$ and $g = -0.40$ for $l_z = \pm 1 \to 0$. Note that nearly the same values are obtained using explicit expressions for the band magnetic dipole moments from Refs. \onlinecite{shi2007,chang1996} (see page 4 of Supplementary information). Thus, unlike absorption, transitions $l_z = 0 \to \pm 1$ prevail in the magneto-optical response at the band edge. The domination of the $\mathcal{A}$ term  is consistent with the experiments, where $\mathrm{Re}/ \mathrm{Im} \ \epsilon_{xy}$ (Fig. \ref{fig:si}b) look similar to derivatives of $\mathrm{Im}/\mathrm{Re} \  \epsilon_{xx}$ (Fig. \ref{fig:si}a). 

To model excitonic effects in silicon we use Eq. (\ref{eq26})  with $\beta=0.2$. This value fulfils the empirical law $\beta=4.615/\epsilon_\infty-0.213$, where $\epsilon_\infty$ is the static dielectric constant, derived for a set of semiconductors with continuum excitons \cite{albrecht1998,botti2004}.
The account of the excitonic effects further improves agreement of the calculated spectra for silicon with the experimental data (Fig. \ref{fig:si}a and b).

It should be noted, however, that though the magnitudes of peaks  in the longitudinal component $\epsilon_{xx}$ of the dielectric tensor agree very well with the experimental results \cite{lautenschlager1987}, the magnitudes of the peaks  in the transverse component $\epsilon_{xy}$ are about a factor of two smaller than in the magneto-optical measurements \cite{agranovich1991}. As discussed above for molecules, the magnitudes of peaks in magneto-optical calculations are strongly dependent on the linewidth assumed. The ratio of the magnitudes of peaks coming from the $\mathcal{A}$ term and those corresponding to the simple absorption scale inversely proportional to the linewidth (see Eq. (\ref{eq29})). Therefore, agreement with the experimental magneto-optical spectra should be improved once the linewidth in the calculations is reduced. Fine-tuning of the linewidth is, however, beyond the scope of the present paper.

In boron nitride (Fig. \ref{fig:bn}), the magneto-optical response of continuum states starts from a prominent peak at the band edge  (Fig. \ref{fig:bn}b). In this material, the first optical transitions take place at the K$^\pm$ points in the corners of the hexagonal Brillouin zone, where phase winding of wavefunctions related to the $C_3$ symmetry imposes coupling to only one light component of the left ($+$) or right ($-$) circular polarization \cite{yao2008,cao2012,xiao2012}. Accordingly, contributions to the magneto-optical spectra from the K$^\pm$ points can be described by a second-order pole (Fig. \ref{fig:bn}c). The map of contributions from different k-points  (Fig. \ref{fig:bn}d) shows that the response is mostly provided by narrow regions in reciprocal space and the sign of the response is opposite in two such regions. Therefore, it can be concluded that the $\mathcal{A}$ term is dominant at the band edge of boron nitride. 

Clearly such a magneto-optical response is related to the valley Zeeman effect \cite{srivastava2015,mitioglu2015,wang2015,macneill2015,li2014}. Since the density of states in two-dimensional materials tends to the Heaviside step function in the limit of zero linewidth, the $\mathcal{A}$ term related to its derivative approaches a delta peak. Thus, discrete peaks in continuum magneto-optical spectra of two-dimensional materials are indicators of the Zeeman splitting.

From the comparison of magneto-optical and optical spectra for boron nitride, we estimate that the change of the magnetic dipole moment upon the excitation at the K$^\pm$ points is  $\Delta m_z^\pm \approx \mp 1.8\mu_\mathrm{B}$. Explicit calculations of the magnetic dipole moments using expressions from Refs. \onlinecite{shi2007,chang1996} give $\mp 0.95\mu_\mathrm{B}$ and $\mp 2.8\mu_\mathrm{B}$ for the valence and conduction bands, respectively, which agrees very well with our estimate. The valley g-factor for the edge of the continuum spectrum according to our calculations is, therefore, $g^\mathrm{vl}=-2\Delta m_z^+/\mu_\mathrm{B} = 3.6$. 

Up to now we have neglected excitonic effects in boron nitride. They, however, are known to be very strong \cite{witz2005}. To describe the first bound exciton in boron nitride we set the parameter $\beta$ in Eq. (\ref{eq26}) at $\beta=17.5$ to reproduce the binding energy of 1.4 eV that follows from the Bethe-Salpeter calculations \cite{witz2005} (Fig. \ref{fig:bn}a).  The absorption (Fig. \ref{fig:bn}a) and magneto-optical  (Fig. \ref{fig:bn}b) spectra computed using this parameter are very similar to those of symmetric molecules like cyclopropane (Fig. \ref{fig:abs}c and d). The valley g-factor deduced from the ratio $\mathrm{Im}\  \epsilon_{xy}/\mathrm{Im}\  \epsilon_{xx}$ at the excitonic peak is about 1.8. It is, therefore, reduced twice compared to the result for the edge of the continuum spectrum. To confirm our estimate, a photoluminescence experiment for boron nitride could be performed by analogy with the measurements for WSe$_2$ (Refs. \onlinecite{srivastava2015,mitioglu2015,wang2015}) and MoSe$_2$ (Refs. \onlinecite{macneill2015,li2014, wang2015}) monolayers (see page 7 of Supplementary information for discussion of g-factors observed for these materials). It should be noted that the qualitative shapes of the  spectra computed with account of the excitonic effects do not depend on the parameter $\beta$ used (see Fig. 3 of Supplementary information) and the valley g-factor changes only by 30\% in the interval of $\beta$ from 10 to 20. 

To summarize, in spite of simplifications made in the present paper for the test calculations, the developed formalism gives realistic results for the magneto-optical response. It provides a unified description of finite and periodic systems and automatically takes into account gauge invariance. Furthermore, it can be straightforwardly extended to the case of higher-order responses to arbitrary electromagnetic fields. 

The efficiency of the implemented procedures for magneto-optics is comparable to standard linear-response calculations of polarizability in the absence of the magnetic field. When local-field effects are included self-consistently, the calculations of magneto-optical spectra for molecules take the same time as polarizability. For solids, the responses at $\pm\Omega_0\pm i\delta$ are needed for magneto-optics as compared only to $\pm\Omega_0+ i\delta$ for simple optics (see the detailed explanation on pages 7 and 8 of Supplementary information) and, therefore, the calculations of magneto-optical spectra take twice as long as those of polarizability. 

\section*{Methods}
The interaction of valence electrons with atomic cores is described using Troullier-Martins norm-conserving pseudopotentials \cite{troullier1991}.  For molecules, the density-averaged self-interaction correction \cite{legrand2002} is applied to avoid spurious transitions to diffuse excited states. The efficient conjugate-gradients solver \cite{jiang2003} is used for the calculation of eigenstates with the tolerance of 10$^{-10}$ and mixing parameter for the Kohn-Sham potential of 0.2 for molecules and 0.1 for solids.  The semiconducting smearing is applied.  
The magnetic gauge correction from Ref. \onlinecite{ismail2001} is added in calculations of magneto-optical spectra of the molecules within the finite-system formulation. The quasi-minimal residual (QMR) method \cite{freund1991} (qmr\_symmetric and qmr\_dotp for the molecules and solids, respectively) with the final tolerance of $10^{-6}$ is used to solve linear equations for projections of derivatives of the density matrix onto unperturbed wavefunctions (Eq. (\ref{eq12})). The local-field effects in the ALDA approximation are taken into account through a self-consistent iteration scheme similar to the ground-state DFT. 

For molecules, the size of the simulation box of 24~\AA~and the spacing of the real-space grid of 0.14~\AA~are sufficient for convergence of the magneto-optical spectra. Only the $\Gamma$ point is used in this case. The geometry of the molecules is optimized till the maximal residual force of 0.01 eV/\AA~using the fast inertial relaxation engine (FIRE) algorithm \cite{bitzek2006}. For boron nitride, we consider the rectangular unit cell of $4.294$ \AA~$\times$ 2.479 \AA~$\times$ 24.0 \AA~with four atoms. For silicon, the cubic unit cell of 5.38 \AA~size with 8 atoms is studied and the grid spacing is increased to 0.25 \AA. Integration over the Brillouin zone is performed according to the Monkhorst-Pack method \cite{monkhorst1976}. Time-reversal and crystal symmetries are taken into account to reduce the number of k-points considered. To take into account time-reversal symmetry, the average of the polarizabilities at frequencies $\Omega$ and $-\Omega$ is computed for irreducible k-points. 3000 irreducible k-points are needed for convergence of the magneto-optical spectra for boron nitride and 6600 for silicon and these are achieved using shifted k-point grids (see the results of calculations using different k-point grids in Figs. 1 and 2 of Supplementary information). 

\subsection*{Data availability}
The datasets generated during the current study are available in the Mendeley Data and NOMAD repositories, http://dx.doi.org/10.17632/749ztg4c9r.1 and  http://dx.doi.org/10.17172/NOMAD/2019.02.13-1, respectively. 

\subsection*{Code availability}
Our implementation is available through the development version of the Octopus code at https://gitlab.com/octopus-code/octopus.git and will be available in future releases at https://octopus-code.org. The code is provided under the 
GNU General Public License. The manual and tutorials can be found at https://octopus-code.org.

\section*{Acknowledgments}
We acknowledge the financial support from the European Research Council (ERC-2015-AdG-694097),  Grupos Consolidados (IT578-13), European Union's H2020 program under  GA no. 646259 (MOSTOPHOS) and no. 676580 (NOMAD) and Spanish Ministry (MINECO) Grant No. FIS2016-79464-P.

\noindent\textbf{Competing interests:} The authors declare no competing interests.

\section*{Author contributions}
A.R. and I.V.T. designed the project. D.A.S. assisted with the Octopus code development. I.V.L. implemented magneto-optical routines, performed the calculations and wrote the manuscript. All the authors discussed the results and commented on the manuscript.

\section*{Additional Information}
\noindent \textbf{Supplementary information} accompanies the paper on the \textit{npj Computational Materials} website ().

\section*{References}
\bibliography{irina_refs}

\end{document}